\documentclass[twocolumn,showpacs,prd]{revtex4}
\usepackage{mathrsfs,bm,multirow}
\usepackage{longtable,lscape}
\usepackage{txfonts}
\usepackage{amssymb}
\usepackage{indentfirst}
\usepackage{graphicx,,booktabs}
\usepackage{epstopdf}
\usepackage{color}
\usepackage{amssymb}

\begin{document}
\title{Hidden-charm decays of $X(3915)$ and $Z(3930)$ as the P-wave charmonia}
\author{Dian-Yong Chen$^{1,2}$}\email{chendy@impcas.ac.cn}
\author{Xiang Liu$^{2,3}$}\email{xiangliu@lzu.edu.cn}
\author{Takayuki Matsuki$^{4,5}$}\email{matsuki@tokyo-kasei.ac.jp}
\affiliation{$^1$Nuclear Theory Group, Institute of Modern Physics, Chinese Academy of Sciences, Lanzhou 730000, China\\
$^2$Research Center for Hadron and CSR Physics, Lanzhou
University $\&$ Institute of Modern Physics of CAS, Lanzhou 730000,
China\\$^3$School of Physical Science and Technology, Lanzhou University, Lanzhou 730000, China
\\
$^4$Tokyo Kasei University, 1-18-1 Kaga, Itabashi, Tokyo 173-8602, Japan\\
$^5$Theoretical Research Division, Nishina Center, RIKEN, Saitama 351-0198, Japan
}
\date{\today}
\begin{abstract}
In this work, we investigate the $X(3915)$ and $Z(3930)$ decays into
$J/\psi\omega$ with the $\chi_{c0}^\prime(2P)$ and $\chi_{c2}^\prime
(2P)$ assignments to $X(3915)$ and $Z(3930)$, respectively. The
results show that the decay width of $Z(3930)\to J/\psi\omega$ is
at least one order smaller than that of $X(3915)\to J/\psi\omega$.
This observation explains why only one structure $X(3915)$ has been
observed in the $J/\psi\omega$ invariant mass spectrum for the
process $\gamma\gamma\to J/\psi\omega $.

\end{abstract}
\pacs{13.25.Gv, 14.40.Rt}

\maketitle


The $\gamma\gamma$ fusion process is an ideal platform to produce the
charmonium-like states. In the past years, the Belle and BaBar experiments
have reported many charmonium-like states in the $\gamma\gamma$ fusion processes.
Among these observations, $X(3915)$ has mass
$M_{X(3915)}=(3915\pm3(\mathrm{stat})\pm 2(\mathrm{syst}))$ MeV and width
$\Gamma_{X(3915)}=(17\pm10(\mathrm{stat})\pm3(\mathrm{syst}))$ MeV. Since $X(3915)$ was
observed in the $J/\psi\omega$ invariant mass spectrum of
$\gamma\gamma\to J/\psi\omega$, the possible quantum number
should be $J^{PC}=0^{++}$ or $J^{PC}=2^{++}$, which
results in the corresponding Belle measurement of
$\Gamma_{X(3915)\to \gamma\gamma}\cdot BR(X(3915)\to
J/\psi\omega) = (61\pm 17(\mathrm{stat})\pm 8(\mathrm{syst}))$ eV or $(18\pm
5(\mathrm{stat})\pm 2(\mathrm{syst}))$ eV \cite{:2009tx}.
As the candidate of charmonium
$\chi_{c2}^\prime(2P)$ ($n^{2s+1}L_J=2^3P_2$), $Z(3930)$ was first observed
in the process $\gamma\gamma\to D\bar{D}$ \cite{Uehara:2005qd}. The
experimental information on $Z(3930)$ gives
$M_{Z(3930)}=3929\pm(\mathrm{stat})5\pm2(\mathrm{syst})$ MeV,
$\Gamma_{Z(3930)}=29\pm10(\mathrm{stat})\pm2(\mathrm{syst})$ MeV, and
$\Gamma_{Z(3930)\to \gamma\gamma}\cdot BR(Z(3930)\to
D\bar{D})=0.18\pm0.05(\mathrm{stat})\pm0.03(\mathrm{syst})$ keV \cite{Uehara:2005qd}.
Later, the BaBar Collaboration also confirmed the observation of
$Z(3930)$ in $\gamma\gamma\to D\bar{D}$ \cite{Aubert:2010ab}.

In Ref. \cite{Liu:2009fe}, the assignments of $X(3915)$ or
$Z(3930)$ as $\chi_{c0}^\prime(2P)$ or
$\chi_{c2}^\prime(2P)$ charmonium states were proposed by
analyzing the mass spectrum and calculating
the strong decay of P-wave charmonium. Later, in Ref. \cite{Lees:2012xs}, the BaBar collaboration announces that the
charmonium like state $X(3915)$ has been confirmed in $\gamma \gamma
\to J/\psi$ process with a spin-parity $J^P=0^+$ \cite{Lees:2012xs},
which is consistent with the prediction in Ref. \cite{Liu:2009fe}.

If $Z(3930)$ is a $\chi_{c2}^\prime(2P)$ state, $Z(3930)$ theoretically
has the hidden charm decay channel $J/\psi\omega$ besides
its observed open charm decay $D\bar{D}$. Hence, the signal of $Z(3930)$
should appear in the same $J/\psi\omega$ invariant mass spectrum as
$X(3915)$ that was observed
by Belle \cite{:2009tx}. However, the experimental data of
the $J/\psi\omega$ invariant mass spectrum shows no
evidence of $Z(3930)$. This fact urges us to
explain why there only exits one signal, $X(3915)$,
observed in the process $\gamma \gamma \to J/\psi\omega$.

In this work, we dedicate ourselves to studying
the $X(3915)$ and $Z(3930)$ decays into $J/\psi\omega$ under the
$\chi_{c0}^\prime(2P)$ and $\chi_{c2}^\prime(2P)$ assignments to $X(3915)$ and $Z(3930)$, respectively.
By this study, we wan to answer whether the decay $Z(3930)\to J/\psi\omega$ is
suppressed compared with $X(3915)\to J/\psi\omega$ under
the P-wave charmonium assignments to $X(3915)$ and $Z(3930)$, which can shed light on the above puzzle.

As higher charmonia, the $X(3915)$ and $Z(3930)$ decays into $J/\psi\omega$ occur via the hadronic loop effects
with the open-charm decay channels as the
intermediate state. This mechanism has been studied in Refs.
\cite{Liu:2006df,Liu:2007ez,Liu:2008yy,Liu:2009iw,Meng:2007fu,
Liu:2006dq,Liu:2009dr,Chen:2009ah} when calculating the hidden-charm
and open-charm decays of charmonium and other charmonium-like states.

\begin{center}
\begin{figure}[htb]
\includegraphics[height=40mm]{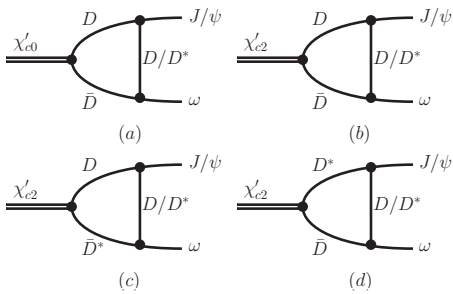}
\caption{The typical diagrams describing the $X(3915)\to J/\psi\omega$
(diagram (a)) and $Z(3930)\to J/\psi\omega$ (diagrams (b)-(d)).
After making the charge conjugate transformation
($D^{(*)}\leftrightarrow\bar {D}^{(*)}$) and the isospin
transformation ($D^{(*)0}\leftrightarrow D^{(*)+}$ and
$\bar{D}^{(*)0} \leftrightarrow D^{(*)-}$), one gets other diagrams.
\label{Fig:gp}}
\end{figure}
\end{center}

{
The discussed $X(3915)$ and $Z(3930)$ are the candidates of the
first radial excitations of $\chi_{c0}$ and $\chi_{c2}$,
respectively. Since the masses of $X(3915)$ and $Z(3930)$ are above
the thresholds of $D\bar{D}$ and $D\bar{D}^*$ and below the
$D^*\bar{D}^*$ threshold, $X(3915)$ and $Z(3930)$ dominantly decay
into $D\bar{D}$ and $D\bar{D}^*$, which contribute the total widths
of $X(3915)$ and $Z(3930)$ (see Ref. \cite{Liu:2009fe} for more
details). As the subordinate decay mode, $J/\psi\omega$ is assumed
from the rescattering contribution of the dominant decays
$X(3915)/Z(3930)\to D\bar{D}, D\bar{D}^*$, which is the reason why
we only consider the intermediate $D\bar{D}$ and $D\bar{D}^*$
contributions in this work.} Under the $\chi_{c0}^\prime(2P)$
assignment to $X(3915)$, the hidden charm decay $X(3915)\to
J/\psi\omega$ occurs through the intermediate states $D\bar{D}$
since $X(3915)$ with $J^{PC}=0^{++}$ dominantly decays into
$D\bar{D}$ as indicated in Ref. \cite{Liu:2009fe}. The hadron level
descriptions of $X(3915)\to D\bar{D}\to J/\psi\omega$ are shown in
Fig. \ref{Fig:gp} (a). The expression for the decay amplitude of
the hidden charm decay $X(3915)\to D\bar{D}\to J/\psi\omega $ reads
as
\begin{eqnarray}
\mathcal{M}[X(3915)\to J/\psi\omega ]=4[\mathcal{A}_{(a)}^{D}+
\mathcal{A}_{(b)}^{D^\ast}].
\label{eq1}%
\end{eqnarray}
As the $\chi_{c2}^\prime(2P)$ state, $Z(3930)$ mainly decays into
$D\bar{D}$ and $D\bar{D}^*+h.c.$  \cite{Liu:2009fe}.
Thus, its hidden charm decay
$Z(3930)\to J/\psi\omega$ is shown in Fig. \ref{Fig:gp} (b)-(d). The amplitude for the processes
$Z(3930)\to D^{(*)}\bar{D}^{(*)}\to J/\psi\omega$ can be expressed as
\begin{eqnarray}
&&\mathcal{M}[Z(3930)\to J/\psi\omega]\nonumber\\
&& =4[\mathcal{M}_{(b)}^{D}+ \mathcal{M}^{D^\ast}_{(b)}+
\mathcal{M}^{D}_{(c)}+\mathcal{M}^{D^\ast}_{(c)}+\mathcal{M}^{D}_{(d)}+
\mathcal{M}^{D^\ast}_{(d)}], \quad \quad
\label{eq2}%
\end{eqnarray}
where the factor 4 in Eq. (\ref{eq1}) and (\ref{eq2}) is resulted
from the charge conjugate and isospin transformations.

To write out the amplitudes corresponding to the diagrams listed
in Fig. \ref{Fig:gp}, we adopt the effective
Lagrangian approach. The effective Lagrangian expressing the interactions of
$X(3915)/Z(3930)$ with $D\bar{D}$ or $D\bar{D}^*+h.c.$ is given by \cite{Colangelo:2003sa}
\begin{eqnarray}
&&\mathcal{L}_{\chi_{cJ}^\prime D^{(\ast)}D^{(\ast)}}=
g_{_{\chi_{c0}^\prime DD}} \chi_{c0}^\prime \mathcal{DD}^\dag
-g_{_{\chi_{c2}^\prime DD}} {\chi_{c2}^\prime}_{\mu\nu}
\partial^{\mu} \mathcal{D} \partial^{\nu}\mathcal{D}^\dag
\nonumber\\ &&\qquad  +ig_{_{\chi_{c2}^\prime D^\ast D}}
\varepsilon_{\mu \nu \alpha \beta}\partial^\mu \chi_{c2}^{\prime \nu
\rho} (\partial^\alpha \mathcal{D}^{\ast \beta}
\partial_{\rho} \mathcal{D}^\dag + \partial^\alpha \mathcal{D}^{\ast \dag \beta}
\partial_{\rho} \mathcal{D}).\quad \quad\label{haha3}
\end{eqnarray}
The couplings of
charmed mesons with the light vector meson $\omega$ or charmonium
$J/\psi$ are constructed in Refs. \cite{Casalbuoni:1996pg,
Oh:2000qr} paying respect to the heavy quark symmetry and the chiral $SU(3)$ symmetry,
and are given below,
\begin{eqnarray}
&&\mathcal{L}_{J/\psi D^{(\ast)}D^{(\ast)}} = i g_{_{J/\psi
\mathcal{D}\mathcal{D}}} \psi_\mu \left(
\partial^\mu \mathcal{D} {\mathcal{D}}^{\dagger} - \mathcal{D}
\partial^\mu {\mathcal{D}}^{\dagger}
\right) \nonumber\\ &&\qquad -g_{_{J/\psi \mathcal{D}^*
\mathcal{D}}}^{} \varepsilon^{\mu\nu\alpha\beta}
\partial_\mu \psi_\nu \left(
\partial_\alpha \mathcal{D}^*_\beta {\mathcal{D}}^{\dagger}
+ \mathcal{D} \partial_\alpha {\mathcal{D}}^{*\dagger}_\beta \right)
\nonumber\\
&&\qquad -i g_{_{J/\psi \mathcal{D}^* \mathcal{D}^*}}^{} \Bigl\{
\psi^\mu \left(
\partial_\mu \mathcal{D}^{*\nu} {\mathcal{D}}_\nu^{*\dagger} -
\mathcal{D}^{*\nu}
\partial_\mu {\mathcal{D}}_\nu^{*\dagger} \right)
\nonumber\\
&&\qquad + \left( \partial_\mu \psi_\nu \mathcal{D}^{*\nu} -
\psi_\nu
\partial_\mu \mathcal{D}^{*\nu} \right) {\mathcal{D}}^{*\mu\dagger}  \mbox{}
\nonumber\\
&& \qquad + \mathcal{D}^{*\mu} \big( \psi^\nu \partial_\mu
{\mathcal{D}}^{*\dagger}_{\nu} - \partial_\mu \psi_\nu
{\mathcal{D}}^{*\nu\dagger} \big) \Bigr\},\\
&& \mathcal{L}_{_{\mathcal{D}^{(\ast)}\mathcal{D}^{(\ast)}
\mathbb{V}}}=-ig_{_{\mathcal{D}\mathcal{D}
\mathbb{V}}}\mathcal{D}_{i}^{\dagger}{
\stackrel{\leftrightarrow}{\partial}}
_{\mu}\mathcal{D}^{j}(\mathbb{V}^{\mu})^{i}_{j} \nonumber\\
&& \quad -2f_{_{\mathcal{D^{*}} \mathcal{D}\mathbb{V}}}
\varepsilon_{\mu\nu\alpha\beta}(\partial^{\mu}\mathbb{V}^{\nu})
^{i}_{j}(\mathcal{D}_{i}^{\dagger}
{\stackrel{\leftrightarrow}{\partial}}^{\alpha}\mathcal{D^{*}}^{\beta
j}  -\mathcal{D^{*}}_{i}^{\beta
\dagger}{\stackrel{\leftrightarrow}{\partial}}^{\alpha}\mathcal{D}^{j})\nonumber\\
&& \quad +
ig_{_{\mathcal{D^{*}}\mathcal{D^{*}}\mathbb{V}}}\mathcal{D^{*}}_{i}^{\nu
\dagger}{\stackrel{\leftrightarrow}{\partial}}_{\mu}\mathcal{D^{*}}_{\nu}^{j}(\mathbb{V}^{\mu})^{i}_{j}
\nonumber\\
&& \quad  +
4if_{_{\mathcal{D^{*}}\mathcal{D^{*}}\mathbb{V}}}\mathcal{D^{*}}_{i\mu}^{\dagger}(\partial^{\mu}\mathbb{V}^{\nu}-\partial^{\nu}
\mathbb{V}^{\mu})^{i}_{j} \mathcal{D^{*}}_{\nu}^{j},\label{ha1}
\end{eqnarray}
where $\mathcal{D}=({D}^0,D^+,D_s^+)$, $(\mathcal{D}^\dag)^T=
(\bar{D}^0,D^-,D_s^-)$ and ${\stackrel{\leftrightarrow}{\partial}}=
{\stackrel{\rightarrow}{\partial}}-{\stackrel{\leftarrow}{\partial}}$.
the light vector nonet meson can form the following $3\times 3$
matrix $\mathbb{V}$,
\begin{eqnarray}
\mathbb{V}=\left(
             \begin{array}{ccc}
               \frac{\rho^0 }{\sqrt{2}}+ \frac{ \omega}{\sqrt{2}} & \rho^+ & K^{\ast +} \\
               \rho^- & \frac{-\rho^0}{\sqrt{2}}+\frac{\omega}{\sqrt{2}} & K^{\ast 0} \\
               K^{\ast -} & \bar{K}^{\ast 0} & \phi \\
             \end{array}
           \right).
\end{eqnarray}

The coupling constants of $\chi_{c0}^\prime\to D\bar{D}$ and
$\chi_{c2}^\prime\to D\bar{D},\,D\bar{D}^*+h.c.$ are obtained by
fitting the total widths of $X(3915)$ and $Z(3930)$, which will be
presented later. While the coupling constants of
$J/\psi$ interacting with a pair of charmed mesons and a coupling
constant of charmed mesons interacting with light vector meson are
given in Table \ref{parameter}
\cite{Casalbuoni:1996pg,Oh:2000qr,Isola:2003fh,Cheng:2004ru}.

\renewcommand{\arraystretch}{1.6}
\begin{table}[htb]
\begin{tabular}{cccccccc}\toprule[1pt]
Coupling& Expression &Value &Coupling& Expression &Value \\\midrule[1pt]
$g_{_{J/\psi
\mathcal{D}\mathcal{D}}}$&$-$&7.71&$g_{_{
\mathcal{D}^*\mathcal{D}^*\mathbb{V}}}$&$\frac{\beta g_V}{\sqrt{2}}$&3.71\\
$g_{_{J/\psi
\mathcal{D}^* \mathcal{D}}}$&$-$&3.98\,GeV$^{-1}$&$f_{_{
\mathcal{D}^*\mathcal{D}^*\mathbb{V}}}$&$\frac{\lambda g_V m_{D^*}}{\sqrt{2}}$&4.64\\
$g_{_{J/\psi
\mathcal{D}^*\mathcal{D}^*}}$&$g_{_{J/\psi
\mathcal{D}\mathcal{D}}}$&7.71&$f_{_{
\mathcal{D}^*\mathcal{D}\mathbb{V}}}$&$\frac{\lambda g_V}{\sqrt{2}}$&2.31\,GeV$^{-1}$\\$g_{_{
\mathcal{D}\mathcal{D}\mathbb{V}}}$&$\frac{\beta g_V}{\sqrt{2}}$&3.71\\
\bottomrule[1pt]
\end{tabular}
\caption{The values of the coupling constants shown in Eqs. (\ref{haha3})-(\ref{ha1}). Here, we take $m_{D}=
(m_{D^0}+m_{D^{\pm}})/2$, $m_{D^*}=(m_{D^{*0}}+m_{D^{*\pm}}) /2$,
$g_V=m_\rho/f_\pi$, $m_\rho=0.77$ MeV, $\beta=0.9$, $\lambda=0.56$
GeV$^{-1}$, $g=0.59$, and $f_\pi=132$ MeV
\cite{Casalbuoni:1996pg,Oh:2000qr,Isola:2003fh,Cheng:2004ru}.
\label{parameter}}
\end{table}

The amplitudes for $\chi_{c0}^\prime(p_0) \to [D(p_1)
\bar{D}(p_2)]D^{(\ast)} (q) \to J/\psi(p_3) \omega(p_4)$
corresponding to Fig. \ref{Fig:gp} (a) are given by
\begin{eqnarray}
&&\mathcal{A}_{(a)}^{D}=(i)^3\int\frac{d^4 q}{(2\pi^4)} \big[
g_{_{\chi_{c0}^\prime \mathcal{DD}}}\big] \big[ig_{_{J/\psi
\mathcal{D}\mathcal{D}}} {\epsilon_{\psi}}_\mu (i q^\mu+i
p_1^\mu)\big] \big[-ig_{_{\mathcal{D}\mathcal{D}\mathbb{V}}}
\nonumber\\&&\quad \quad \times (-i
p_{2\nu}+iq_\nu){\epsilon_\omega}^\nu\big] \frac{i}{p_1^2-m_{D}^2}
\frac{i}{p_2^2-m_{D}^2}\frac{i}{q^2-m_{D}^2}\,\mathcal{F}^2 (q^2),
\nonumber\\
&&\mathcal{A}^{D^\ast}_{(a)}=(i)^3\int\frac{d^4 q}{(2\pi^4)}\big[
g_{_{\chi_{c0}^\prime \mathcal{DD}}} \big] \big[-g_{_{J/\psi
\mathcal{D}^*\mathcal{D}}}\varepsilon^{\mu\nu\alpha\beta}(ip_{3\mu}){\epsilon_{\psi}}_\nu
(i q_\alpha)\big]  \nonumber\\&&\quad \quad\times \big
[-2f_{_{\mathcal{D}^*\mathcal{D}\mathbb{V}}}\varepsilon_{\sigma\lambda\rho\xi}(ip_4^\sigma)\epsilon_{\omega}^\lambda
(ip_2^\rho-iq^\rho)\big]\frac{i}{p_1^2-m_{D}^2} \nonumber\\&& \quad
\quad\times
\frac{i}{p_2^2-m_{D}^2}\frac{i\tilde{g}_\beta^{\xi}(q)}{q^2-m_{D^*}^2}\mathcal{F}^2
(q^2), \label{Eq:M0}
\end{eqnarray}
where $\mathcal{A}_{(a)}^{D}$ and $\mathcal{A}_{(a)}^{D^\ast}$ are
the amplitudes corresponding to diagram (a) in Fig. \ref{Fig:gp}
with the $D$ and $D^\ast $ meson exchanges, respectively. Similarly, we
can easily write out the the expressions of the decay amplitudes for $Z(3930)\to J/\psi \omega$ corresponding to
Fig. \ref{Fig:gp} (b)-(d), which are

\begin{eqnarray}
&&\mathcal{M}^{D}_{(b)} = (i)^3\int\frac{d^4
q}{(2\pi^4)}\big[-g_{_{\chi_{c2}^\prime \mathcal{DD}}}
\epsilon_{\chi_{c2}^\prime}^{\mu \nu}(ip_{1\mu})(ip_{2\nu}) \big] \nonumber\\
&&\quad \quad \times \big[ig_{_{J/\psi
\mathcal{D}\mathcal{D}}}{\epsilon_{\psi}}_\rho (i q^\rho+i p_1^\rho)
\big] \big[-ig_{_{\mathcal{D}\mathcal{D}\omega}}(-i
p_{2\tau}+iq_\tau) \epsilon_\omega^\tau\big]\nonumber\\&& \quad
\quad \times \frac{i}{p_1^2-m_{D}^2}
\frac{i}{p_2^2-m_{D}^2}\frac{i}{q^2-m_{D}^2}\,\mathcal{F}^2(q^2),\nonumber\\\label{Eq:M1}
&&\mathcal{M}^{D^\ast}_{(b)}=(i)^3\int\frac{d^4
q}{(2\pi^4)}\big[-g_{_{\chi_{c2}^\prime \mathcal{DD}}}
\epsilon_{\chi_{c2}^\prime}^{\mu \nu}(ip_{1\mu})(ip_{2\nu}) \big]
  \big[-g_{_{J/\psi
\mathcal{D}^*\mathcal{D}}}\nonumber\\ && \quad \quad \times
\varepsilon_{\theta \rho \alpha
\beta}(ip_{3}^{\theta})\epsilon_{\psi}^\rho (i q^\alpha)
\big]\big[-2f_{_{\mathcal{D}^*\mathcal{D}\omega}}\varepsilon_{
\sigma\tau \lambda \phi}  (ip_4^\sigma) \epsilon_{\omega}^\tau
(ip_2^\lambda-iq^\lambda)\big]\nonumber\\&&\quad \quad \times
\frac{i}{p_1^2-m_{D}^2} \frac{i}{p_2^2-m_{D}^2}\frac{i
\tilde{g}^{\beta \phi}(q)}{q^2-m_{D^*}^2}
\mathcal{F}^2(q^2),\nonumber\\
&&\mathcal{M}^{D}_{(c)}=(i)^3\int\frac{d^4 q}{(2\pi^4)} \big[
ig_{\chi_{c2}^\prime D^\ast D} \varepsilon_{\delta \mu \theta \phi}
(-ip_0^\delta) \epsilon_{\chi_{c2}^\prime}^{\mu \nu} (ip_2^{\theta})
  (ip_{1 \nu}) \big]  \nonumber\\ && \quad \quad \times \big[ig_{_{J/\psi
\mathcal{D}\mathcal{D}}}\epsilon_{\psi}^\rho (i q_\rho+i p_{1\rho})
\big] \big[-2f_{_{\mathcal{D}^*\mathcal{D}
\omega}}\varepsilon_{\sigma\tau\lambda\alpha} (ip_4^\sigma)
\epsilon_{\omega}^\tau   \nonumber\\&& \quad \quad \times
(-ip_2^\lambda +iq^\lambda)\big] \frac{i}{p_1^2-m_{D}^2}
\frac{i\tilde{g}^{\phi\alpha}(p_2)}{p_2^2- m_{D^*}^2}
\frac{i}{q^2-m_{D}^2}\,\mathcal{F}^2(q^2),\nonumber\\
&&\mathcal{M}^{D^\ast}_{(c)}=(i)^3\int\frac{d^4 q}{(2\pi^4)} \big[
ig_{\chi_{c2}^\prime D^\ast D} \varepsilon_{\delta \mu \theta \phi}
(-ip_0^\delta) \epsilon_{\chi_{c2}^\prime}^{\mu \nu} (ip_2^{\theta})
(ip_{1 \nu}) \big] \nonumber\\&& \quad \quad \times
\big[-g_{_{J/\psi \mathcal{D}^*\mathcal{D}}}\varepsilon_{\lambda
\rho \alpha \beta} (ip_{3}^{\lambda}) \epsilon_{\psi}^\rho (i
q^\alpha) \big] \big [ig_{_{\mathcal{D}^*\mathcal{D}^*\omega}}
(-ip_{2\tau}+i q_{\tau}) \nonumber\\&& \quad \quad \times
\epsilon_{\omega}^\tau g_{\zeta \sigma }
 +4if_{_{\mathcal{D}^*\mathcal{D}^*\omega}}
\epsilon_{\omega}^{\tau}\big(ip_{4 \zeta} g_{\sigma \tau} - i
p_{4\sigma} g_{\tau \zeta} \big)\big]\frac{i}{p_1^2-m_{D}^2}
\nonumber\\&&\quad \quad \times \frac{i\tilde{g}^{\phi
\sigma}(p_2)}{p_2^2-m_{D^*}^2} \frac{i \tilde{g}^{\zeta \beta}(q)}{
q^2-m_{D^*}^2}\,\mathcal{F}^2(q^2),\nonumber \\
&&\mathcal{M}^{D}_{(d)}=(i)^3\int\frac{d^4 q}{(2\pi^4)} \big[
ig_{\chi_{c2}^\prime D^\ast D} \varepsilon_{\delta \mu \theta \phi}
(-ip_0^\delta) \epsilon_{\chi_{c2}^\prime}^{\mu \nu} (ip_1^\theta)
 (ip_2^\nu) \big] \nonumber\\&&\quad \quad  \times  \big[-g_{_{J/\psi D^\ast D}}
\varepsilon_{\lambda \rho \alpha \beta } (ip_3^\lambda)
\epsilon_{\psi}^\rho (-ip_1^\alpha) \big] \big[ ig_{_{DD\omega}}
\epsilon_{\omega}^\tau (ip_{2\tau} -iq_{\tau})\big]
\nonumber\\&&\quad \quad \times\frac{i\tilde{g}^{\phi
\beta}(p_1)}{p_1^2-m_{D^\ast}^2} \frac{i}{p_2^2-m_{D}^2}
\frac{i}{q^2-m_{D}^2}\,\mathcal{F}^2(q^2),\nonumber\\
&&\mathcal{M}^{D^\ast}_{(d)} =(i)^3\int\frac{d^4 q}{(2\pi^4)} \big[
ig_{\chi_{c2}^\prime D^\ast D} \varepsilon_{\delta \mu \theta \phi}
(-ip_0^\delta) \epsilon_{\chi_{c2}^\prime}^{\mu \nu}
(ip_1^\theta)(ip_2^\nu) \big]  \nonumber\\&& \quad \quad \times
\big[ -ig_{_{J/\psi D^\ast D^\ast}} \epsilon_{\psi}^\rho \big(
g_{\alpha \beta} (-ip_{2\rho}+iq_{\rho})  + g_{\beta \rho}
(ip_{3\alpha} + ip_{1\alpha} ) \nonumber\\&& \quad \quad + g_{\alpha
\rho} (-iq_{\beta} -ip_{3 \beta}) \big) \big] \big[ -2f_{_{D^\ast D
\omega}} \varepsilon_{\sigma \tau \lambda \zeta} (ip_4^\sigma)
\epsilon_{\omega}^\tau (-iq^\lambda \nonumber\\&&\quad \quad
+ip_2^\lambda) \big] \frac{i\tilde{g}^{\phi
\beta}(p_1)}{p_1^2-m_{D^\ast}^2} \frac{i }{ p_2^2-m_{D}^2}
\frac{i\tilde{g}^{\alpha
\zeta}(q)}{q^2-m_{D^*}^2}\,\mathcal{F}^2(q^2).  \label{Eq:M6}
\end{eqnarray}
with $\tilde{g}^{\alpha \beta}(p)=-g^{\alpha \beta}+p^\alpha
p^\beta/m_{D^\ast}^2$, where $\mathcal{F}(q^2)$ is the form factor, which
is introduced not only to compensate the off-shell effects of the
charmed meson but also to describe the structure effects of the
vertex of a charmed meson pair interacting with $J/\psi$ or
$\omega$. In this work, we adopt the form factor in the form ,
\begin{eqnarray}
\mathcal{F}(q^2)=\left( {m_{E}^2-
\Lambda^{2}}\over{q^2- \Lambda^{2}}\right)^N, \ \ \
\left\{
  \begin{array}{ll}
    N=1, & \hbox{Monopole form;} \\
    N=2, & \hbox{Dipole form.}
  \end{array}
\right.
\end{eqnarray}
where $q$ and $m_{E}$ are the momentum and the mass of the exchanged charmed meson, respectively. Furthermore,
$\Lambda$ can be parameterized as $\Lambda=m_{E}+\alpha
\Lambda_{QCD}$ with a dimensionless parameter
$\alpha$ and $\Lambda_{QCD}=220$ MeV.
The parameter
$\alpha$ is of order unity and depends on the specific process
\cite{Cheng:2004ru,Chen:2009ah}.

With the above elaborate expressions for the amplitudes, one can
obtain the partial decay width for $\chi_{cJ}^\prime \to J/\psi
\omega, (J=0,2)$ as,
\begin{eqnarray}
d\Gamma_{\chi_{cJ}^\prime \to J/\psi \omega} =\frac{1}{2J+1}
\frac{1}{32 \pi^2} \overline{\left|\mathcal{M}_{\chi_{cJ}^\prime \to
J/\psi \omega} \right|^2} \frac{|\vec{p}|}{m_{\chi_{cJ}^\prime}^2}
d\Omega,
\end{eqnarray}
where the overline indicates the sum over the polarizations of vector
meson $J/\psi, \omega$ and tensor meson $\chi_{c2}^\prime$, and
$\vec{p}$ indicates the three momentum of $J/\psi$ in the initial
state at rest.

\begin{center}
\begin{figure}[htb]
\includegraphics[height=65mm]{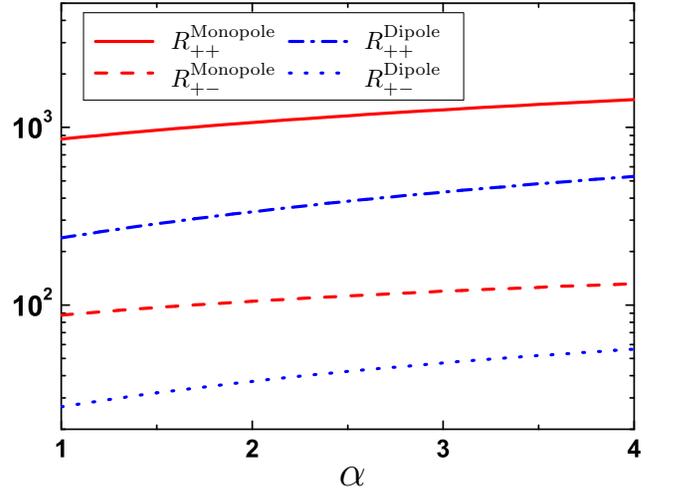}
\caption{(Color online.) $\alpha$ dependence of
the ratio of the width of $X(3915)\to
J/\psi\omega$ to that of $Z(3930)\to J/\psi\omega$. Here, we define $R_{++}=\Gamma[X(3915)\to
J/\psi\omega]/\Gamma[Z(3930)\to J/\psi\omega]_{++}$ and $R_{+-}=\Gamma[X(3915)\to
J/\psi\omega]/\Gamma[Z(3930)\to J/\psi\omega]_{+-}$. In addition,
we use subscripts "Monopole" and "Dipole" to distinguish different results by taking monopole and dipole form factors in the calculation, respectively.
The calculated results are $R_{++}^{\mathrm{Monopole}}=850\sim 1400$,
$R_{++}^{\mathrm{Dipole}}=239 \sim 558$, $R_{+-}^{\mathrm{Monopole}}=87\sim 130$ and  $R_{+-}^{{\mathrm{Dipole}}}=27 \sim 59$.} \label{Fig:x3915_0+}
\end{figure}
\end{center}

If $X(3915)$ is a $\chi_{c0}^\prime(2P)$ state, $D\bar{D}$ is its
dominant decay. Hence, we can use the experimental width of $X(3915)$
\cite{:2009tx} to determine the coupling constant of
$\chi_{c0}^\prime \to D\bar{D}$ interaction, i.e.,
$g_{\chi_{c0}^{\prime}D\bar{D}}= 2.37$ GeV. However, for $Z(3930)$,
there exist two main decay modes $D\bar{D}$ and $D\bar{D}^*+h.c.$.
Since at present experiments did not give the ratio of
$BR(Z(3930)\to D\bar{D})$ to $BR(Z(3930)\to D\bar{D}^*+h.c.)$,
we must determine these corresponding coupling constants from the
theoretical results estimated by quark pair creation model.  
In Ref. \cite{Liu:2009fe}, the wave functions of $\chi_{cJ}^\prime$ are simulated by a simple harmonic oscillator wave function with a parameter $R$, which maens a root mean square radius of the wave function. The partial and total decay widths of $\chi_{cJ}^\prime$ are dependent on this unique parameter $R$.
One can determine the parameter value $R \simeq 1.9 \mathrm{GeV}^{-1}$ in the spatial wave function from the partial decay width of
$\chi_{c0}^\prime$ under the assumption $\Gamma_{\chi_{c0}^\prime
\to D\bar{D}} \simeq \Gamma_{\chi_{c0}^\prime}^{\mathrm{tot}}$. With
the parameter $R$ estimated by the center value of
$\Gamma_{\chi_{c0}^\prime}^{\mathrm{tot}}$, we obtain
$|g_{\chi_{c2}^\prime D D}|=11.69$ $\mathrm{GeV}^{-1}$ and
$|g_{\chi_{c2}^\prime D^\ast D}|= 7.83$ $\mathrm{GeV}^{-2}$.

For $Z(3930)$ with the assignment of $\chi_{c2}^\prime$, it
dominantly decays into $D\bar{D}$ and $D^\ast \bar{D}+h.c $. The
absolute values of the coupling constants between $\chi_{c2}^\prime$
and charmed meson pairs are evaluated by the quark pair creation model.
However, the relative sign of coupling constants
$g_{\chi_{c2}^\prime DD}$ and $g_{\chi_{c2}^\prime D^*D}$ in Eq.
(\ref{haha3}) can be either positive or negative, which corresponds to the subscripts $++$ and
$+-$ shown in Fig. \ref{Fig:x3915_0+}, respectively.
Thus, we discuss two cases for
$Z(3930)\to J/\psi\omega$.

In Fig. \ref{Fig:x3915_0+}, we give the ratio of the width of
$X(3915)\to J/\psi\omega$ to that of $Z(3930)\to J/\psi\omega$. This
result shows that the width of $X(3915) \to J/\psi \omega$ is at
least one order of magnitude larger than that of $Z(3930)\to J/\psi
\omega$ in two different cases (see Fig. \ref{Fig:x3915_0+} for more
details). Although the decay width for $\chi_{c0}^\prime
/\chi_{c2}^\prime \to J/\psi \omega$ calculated in the this work
strongly depends on the parameter $\alpha$, the ratio of
the width of $X(3915)\to J/\psi\omega$ to that of $Z(3930)\to
J/\psi\omega$ has very large value and is weakly dependent on the
parameter $\alpha$ as shown in Fig. \ref{Fig:x3915_0+}. Such a large
ratio can explain why Belle only reported one enhancement
structure $X(3915)$ in the $J/\psi\omega$ invariant mass spectrum of
the $\gamma\gamma\to J/\psi\omega$ process.

\begin{figure}
\centering
\includegraphics[height=65mm]{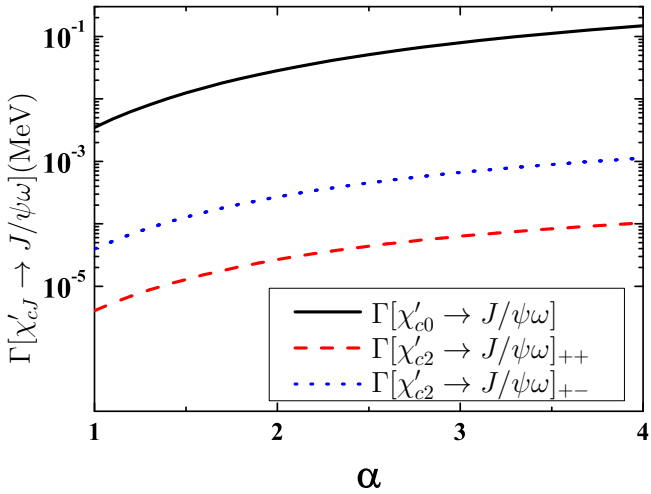}
\caption{$\alpha$ dependence of the partial decay widths of $\chi_{cJ}^\prime \to J/\psi \omega$, where $X(3915)$ and $Z(3930)$ are assigned as $\chi_{c0}^\prime $ and $\chi_{c2}^\prime$, respectively. The loop integrals in Eqs. (\ref{Eq:M0}) and (\ref{Eq:M6})  are evaluated by the Cutkosky cutting rules. The subscripts $++$ and $+-$ are the same as those in Fig .\ref{Fig:x3915_0+}. \label{Fig:width}}
\end{figure}

In addition, the $\alpha$ dependence of the $\chi_{cJ}^\prime \to J/\psi \omega$ partial decay widths is presented in Fig. \ref{Fig:width}. Here, we take the monopole form factor as an example. As one find in Fig. \ref{Fig:width}, the partial decay widths are strongly dependent on the parameter $\alpha$. As for $\chi_{c0}^\prime \to J/\psi \omega$, the partial decay width varies from $3.5 \times 10^{-3}~\mathrm{MeV}$ to 0.15 MeV in the range $1<\alpha <4$. While for $\chi_{c2}^\prime$, the partial decay width of $\chi_{c2}^\prime \to J/\psi \omega$ varies from $4.1 \times 10^{-6}$ MeV to $1.1 \times 10^{-4}$ MeV or $4.0 \times 10^{-5}$ MeV to $1.1 \times 10^{-3}$ MeV depending on the relative sign between $g_{\chi_{c2}^\prime DD}$ and $g_{\chi_{c2}^\prime D^\ast D}$. In the present calculations, the loop integrals in Eqs. (\ref{Eq:M0}) and (\ref{Eq:M6}) are evaluated by the Cutkosky cutting rules, where only the imaginary part of the amplitudes are considered.
As for the case of the dipole form factor, the magnitudes of the partial decay widths are at least one order smaller than the corresponding ones estimated by the monopole form factor. The calculations in Ref. \cite{Chen:2012nva} also indicate that the monopole form factor is more suitable to estimate the partial decay widths of $\chi_{cJ}^\prime \to J/\psi \omega$ process.

As indicated above, the absolute decay widths of $X(3915)\to
J/\psi\omega$ and $Z(3930)\to J/\psi\omega$ are strongly dependent on
parameter $\alpha$, which makes that there exist uncertainty of the
prediction of these decay widths. In addition, we notice that
extracting the decay widths of $X(3915)\to J/\psi\omega$ to that of
$Z(3930)\to J/\psi\omega$ via the experimental data depends on our
understanding of two photon decay width of $X(3915)$ and $Z(3930)$,
where its predicted two photon decay width varies with different
models. In Ref. \cite{Godfrey:1985xj, Munz:1996hb,
Ebert:2003mu,Hwang:2010iq}, the two-photon decay width is about
$1\sim 2$ keV in the relativistic quark model. While the Salpeter
method indicates that the decay width for $\chi_{c0}^\prime$ can be
larger than 3 keV in a relativistic form and about 5.47 keV in a
non-relativistic form \cite{Wang:2007nb}.
If the center values of the total decay width of $\chi_{c0}^\prime$ and the measured branching ratio $\Gamma_{\chi_{c0}^\prime \to \gamma \gamma }\mathcal{B}(\chi_{c0}^\prime \to J/\psi \omega)$ are adopted, the partial decay width of  $\chi_{c0}^\prime \to J/\psi \omega$ can be less than two hundred keV to 1 MeV depending on a different choice of $\Gamma_{\chi_{c0}^\prime \to \gamma \gamma}$. In the present work, the evaluated partial decay width can reach to 150 keV for $\alpha =4$, which is consistent with the experimental measurements \cite{:2009tx}.

In summary, $X(3915)$ reported by the Belle Collaboration is the second
enhancement observed in the $\gamma\gamma$ fusion process. As
indicated in Ref. \cite{Liu:2009fe}, $X(3915)$ is a good candidate
of $\chi_{c0}^\prime(2P)$, i.e., the first radial excitation of
$\chi_{c0}(3414)$. Besides its open charm decay, study of the
hidden charm decay of $X(3915)$ will provide a key hint to
understand the properties of $X(3915)$ and further test P-wave
charmonium explanation to $X(3915)$ in Ref. \cite{Liu:2009fe}. Since
the mass of $X(3915)$ is above the threshold of $D\bar D$ and
dominantly decays into $D\bar{D}$, hadronic loop
effects
\cite{Liu:2006df,Liu:2007ez,Liu:2008yy,Liu:2009iw,Meng:2007fu,
Liu:2006dq,Liu:2009dr,Chen:2009ah} will play an important role to the
hidden charm decay $X(3915)\to J/\psi\omega$, which in fact is
resulted from the coupled channel effects. In this work, we have performed
the calculation of the $X(3915)\to J/\psi\omega$ processes.

Before the observation of $X(3915)$, Belle once reported a state
named as $Z(3930)$ in the $\gamma\gamma$ fusion
\cite{Uehara:2005qd,Aubert:2010ab}, which is also a P-wave
charmonium state of the first radial excitation. $Z(3930)$ should
decay into $J/\psi\omega$, which seems to indicate that there should
exist two peaks being close to each other in the $J/\psi\omega$
invariant mass spectrum given by Belle \cite{:2009tx}. However,
presently only one structure corresponding to $X(3915)$ was observed
\cite{:2009tx}. In order to explain this contradiction, in this work
we have further studied $Z(3930)\to J/\psi\omega$ by the
intermediate states $D\bar{D}$ and $D\bar{D}^*+h.c.$. The results
illustrated in Fig. \ref{Fig:x3915_0+} show that the partial decay
width of $Z(3930)\to J/\psi\omega$ is suppressed when compared with
that of $X(3915)\to J/\psi\omega$, which explains why $Z(3930)$ can
not be observed in the $J/\psi\omega$ invariant mass spectrum.

As more charmonium-like states are observed in the $\gamma\gamma$
fusion process
\cite{:2009tx,Uehara:2005qd,Aubert:2010ab,Shen:2009vs}, they provide
us a better chance to explore the property of these states,
especially P-wave charmonium states \cite{Liu:2009fe}. The study of
the hidden charm decay of $X(3915)$ in this work supports the
proposal of $\chi_{c0}^\prime(2P)$ assignment to $X(3915)$ in Ref.
\cite{Liu:2009fe}. Besides applying the hidden charm and the open
charm decays of $X(3915)$ to test the $\chi_{c0}^\prime$ assignment
to $X(3915)$, we suggest that the angular distribution analysis of
$X(3915)$ in future experiment will be valuable to test the
$\chi_{c0}^\prime(2P)$ explanation to $X(3915)$ since the $J^{PC}$
quantum number of $X(3915)$ must be $0^{++}$. Although $Z(3930)$ is
well established as $\chi_{c2}^\prime(2P)$ state
\cite{Uehara:2005qd,Aubert:2010ab}, its hidden charm decay behavior
is unclear before this work. Performing the calculation of
$Z(3930)\to J/\psi\omega$ by the hadronic loop mechanism, we further
learn that the branching ratio of $Z(3930)\to J/\psi\omega$ is at
least one order smaller than that of $X(3915)\to J/\psi\omega$,
which not only successfully explains only one enhancement $X(3915)$
appearing in the $J/\psi\omega$ invariant mass spectrum but also
tests the hadronic loop effects which is an important
non-perturbative mechanism on the decays of charmonium or
charmonium-like state
\cite{Liu:2006df,Liu:2007ez,Liu:2008yy,Liu:2009iw,Meng:2007fu,
Liu:2006dq,Liu:2009dr,Chen:2009ah}.

\vfil

\noindent {\bf Acknowledgements}: This project is supported by the
National Natural Science Foundation of China under Grant Nos.
11175073, 11005129, 11375240, 11035006, the Ministry of Education of
China (FANEDD under Grant No. 200924, DPFIHE under Grant No.
20090211120029, NCET, the Fundamental Research Funds for the Central
Universities, the Fok Ying Tung Education Foundation (No. 131006),
and the West Doctoral Project of Chinese Academy of Sciences.

\end{document}